\documentclass[11pt,twoside]{article}

\usepackage{asp2006}
\usepackage{epsf}
\usepackage{lscape}
\usepackage{graphicx}
\usepackage{subfigure}

\begin{document}
\title{Dynamics and origins of the young stars in the Galactic center}   
\author{Hagai B. Perets}   
\affil{Harvard-Smithsonian Center for Astrophysics, 60 Garden St. Cambirdge MA, USA 02138}    

\begin{abstract}
The environment near the massive black hole (MBH) in the Galactic
center is very hostile for star formation. Nevertheless, many young
stars (both O and B stars) are observed close the MBH. The B-stars
seems to have an isotropic, continuous distribution between 0.01 pc
and up to a pc. The O stars, in contrast, seem to be distributed in
a coherent disk like configuration, extending only between $\sim0.04$
pc to $\sim0.5$ pc. Our current understanding favors an in-situ formation
origin for the more massive (O and Wolf-Rayet) stars, in gaseous disk
and/or streams from an infalling gas clump. The B-stars seem to have
a different origin, more likely through a dynamical capture, following
binary disruption by the MBH. This scenario could also be able to
explain the origin of hypervelocity stars in the Galactic halo. These
and other possible origins of the young stars in the Galactic center
are briefly reviewed and their possible observational signatures and
constraints are detailed.
\end{abstract}

\vspace{-0.3cm}

\section{Introduction}

High resolution observations have revealed the existence of many apparently
normal young OB (including Wolf-Rayet; WR) stars in the galactic center
(GC), where tidal forces exerted by the massive black hole (MBH; \citep{sch+02b,ghe+03a})
are likely to inhibit regular star formation in regular molecular
clouds. The existence and properties of such stars could give us important
clues for understanding of the GC environment \citep[see][for a review]{ale05}.
They could also help constrain the origin and evolution of hypervelocity
stars observed in the Galactic outskirts \citep{hil88,bro+06a,per09},
which in principle could probe the potential and dark matter component
of the Galaxy \citep{gne+05,yuq+07,per+09}. 

The young stars observed in the central pc near the MBH, could be
divided into two seemingly distinct stellar populations, which differ
both in their types (B-stars vs. O or WR stars) and in their kinematic
properties. The young B-stars population ($\sim7-15\, M_{\odot}$;
fainter stars can not currently be resolved) includes a few tens of
stars with an isotropic distribution extending from $\sim0.01$ pc
all through the central pc \citep{bar+10}. For some of these stars,
in the central 0.04 pc (so called the 'S-stars' or the 'S-cluster'),
full orbital solutions are known, showing them to have relatively
high eccentricities ($0.3\le e\le0.95$) \citep{ghe+03a,eis+05},
with approximately thermal eccentricity distribution, and random orbital
orientations. Although the full kinematic properties of the B-stars
outside this region are not known, the available data suggest their
distribution is isotropic with similarly relaxed eccentricity distribution. 

The other young stars (mostly O and WR stars) reside in a more restricted
region, between 0.04 pc to 0.5 pc, in seemingly two coherent structures.
Most of these reside in a stellar disk moving clockwise in the gravitational
potential of the MBH \citep{lev+03,gen+03a,lu+06,pau+06,tan+06,bar+09}.
The second structure is less coherent and its exact nature (and existence)
is still unknown and debated. The orbits of the stars in the CW disk
have an average eccentricity of $\sim0.35$ and the opening of the
disk is $h/R\sim0.1$, where $h$ is the disk height and $R$ is its
radius. The disk structure is warped at large angles ($65^{\circ}$).
Most of the stars outside the CW disk reside in somewhat less coherent
structure, between 0.3-0.5 pc from the MBH, and highly inclined with
respect to the CW disk. A small fraction of the young O stars do not
reside in either of these structures at some intermediate inclinations.
The current knowledge on the observed properties of the young stars
in the GC are discussed in several recent papers (see e.g. \citealp{bar+10}). In table 1, we summarize the observed properties
of the young stars in the GC, which should be satisfied by models
of their origin and evolution.

{\scriptsize }%
\begin{table}
{\scriptsize }\begin{tabular}{|l|l|l|l|l|}
\hline 
{\tiny Stellar} & {\tiny Masses and } & {\tiny Morphology} & {\tiny Radial ($n(r)$)} & {\tiny Eccentricity}\tabularnewline
{\tiny Type} & {\tiny Lifetime/age} &  & {\tiny Distribution} & {\tiny Distribution}\tabularnewline
\hline
\hline 
{\tiny B} & {\tiny 7-15 $M_{\odot}$ } & {\tiny Isotropic} & {\tiny $r^{-1.1}$} & {\tiny $\sim$Thermal}\tabularnewline
 & {\tiny 10-50 Myrs} & {\tiny (spherical)} &  & {\tiny $\left\langle e\right\rangle \sim0.7$}\tabularnewline
 & {\tiny regular IMF} &  &  & \tabularnewline
\hline 
{\tiny O and WR} & {\tiny 15-60 $M_{\odot}$;} & {\tiny Clockwise warped ($65^{\circ}$) disk } & {\tiny $r^{-2}$} & {\tiny $\left\langle e\right\rangle \sim0.35$}\tabularnewline
 & {\tiny{} 4-6 Myrs} & {\tiny ($H/R\sim0.1$) + coherent highly } & {\tiny (in the CW disk)} & {\tiny (in the CW disk)}\tabularnewline
 & {\tiny top heavy IMF} & {\tiny inclined disk/stream structure} &  & \tabularnewline
\hline
\end{tabular}{\scriptsize \par}

{\scriptsize \caption{{\scriptsize Properties of the GC young stars}}
}
\end{table}
{\scriptsize \par}

In the following we briefly overview suggested scenarios for the origin
of the young stars in the Galactic center. Several models suggested
these stars to be older stars, which only appear to be young. However,
current observations show the young GC stars are apparently normal,
genuinely young, and massive stars. We therefore focus on other more
favorable scenarios, producing \emph{young} stars (see \citet{ale05}
for an overview of the young stars impostors scenario, as well as
some of less recent literature on some of the other models).

\section{Origins and evolution}

\subsection{In-situ star-formation from infalling gaseous clumps}

It was suggested that the young OB stars in the GC were formed in-situ
in the central pc a few million years ago in gaseous disks and/or
streams formed from infalling and/or colliding gaseous clumps \citep{mor93}.
Analytic calculations and simulations \citep{nay+05b,lev07,bon+08,hob+09}
have shown that stars could form in such fragmenting clumps to form
stellar disks and/or other coherent structures, in the region of a
few 0.01 pcs up to a few 0.1 pcs from the MBH. These could, in principle,
be consistent with the observed young stellar structures (disk and
a secondary inclined structure). This scenario could also explain
the radial distribution of OB stars in the disk and possibly also
explain their eccentricities. Moreover, some of the models studied
produce a top heavy mass function (MF) for the newly formed stars,
possibly consistent with observations \citep{bar+10}. 

Note, however, that our poor understanding of the initial conditions
and the star formation processes in the GC, allows for a wide parameter
space to play with, which, naturally, raise difficulties in constraining
(or strictly falsifying) such models. Nevertheless, the robustness
of producing some sort of star formation in the GC region under a
variety of conditions explored in the literature suggest these models
as the currently most promising scenarios for the origin of the young
stellar disk and stellar structures (although less likely the origin
of the isotropic B-stars population, as discussed below).

\subsection{In-situ star-formation followed by rapid migration}

Non of the models for in-situ star formation in the central pc suggests
the formation of stars as close to the MBH as the S-stars, or the
apparent existence of two distinct young O and B stellar populations.
Producing both the disk and isotropic B-stars population in the same
scenario (and in particular the inner S-stars) requires a somewhat
fine-tuned scenario. One requires a selective process which works
differentially on stars of different masses. In any case, an additional
process would be required for the migration of stars from the outer
to inner regions in the central pc in order to produce the S-stars
close to the MBH. 

Two-body relaxation processes works, in principle, differentially
on stars of different masses, through mass segregation (energy equipartition)
processes. For example, such processes in an isolated stellar disk
would somewhat segregate the more massive stars into a thinner disk
\citep{ale+07,per+08b} producing mass stratified populations. However,
this effect is relatively small. Moreover, the most important component
for relaxation in the GC is likely to be the stellar black holes population
of the stellar cusp, which work through resonant relaxation \citep{per+08b,loc+09,per+09b}.
Segregation into two different stellar population of different masses
regimes is not likely to insue in this case. Moreover, such scattering
of stars could not produce the population of S-stars closer to the
MBH (\citealp{per+08b}; also Perets et al., in prep.). Encounter
of binary-single stars \citep{cua+08,per+08c} can also have only
a small effect in producing the isotropic B-stars population from
a thin stellar disk. Perturbation by massive perturbers \citep{per+07}
such as infalling IMBHs \citep{yuq+07b,gua+09} or other stellar disks
\citep[; also Gualandris et al., in prep.]{loc+09} do not differentiate
between stars of different masses either.

Given the lacking suggestion for a mass differentiating process, a
different route can be taken. We can suggest that two distinct epochs
of in-situ star-formation have occurred. Even two such epochs would
still require a migration process of some of the stars from the disk
into the inner regions close to the MBH, where stars do not form in-situ.
One should mention, in this context, that in some cases different
mass function was found for stars formed at different structures in
the same simulation \citep{hob+09}. We can therefore either suggest
two epochs of star formation happening at different times, or a single
epoch producing two distinct population. In either case the rapid
migration producing the S-stars should affect only one of the stellar
populations formed. 

Two distinct star formation epochs could naturally produce more massive
and less massive stellar populations. Even if both epochs produced
stellar population with the same initial mass function, the most massive
stars from the first epoch may already end their life, leaving a stellar
population of less massive stars. However, comparing the top heavy
mass function of the disk stars, as suggested from current observation,
with the regular mass function of the isotropic B-stars disfavor this
scenario (otherwise much more B-stars should have been observed in
the stellar disk).

This bring us to an interesting very general requirement applicable
to any scenario in which the S-stars formed far from their current
positions. We can term this the model efficiency. 
In any such scenario only a fraction, $f_{mig}$, of the
stars formed in some external region (e.g. the stellar disk, the central
pc, or even stars outside the central pc) migrate to become S-stars.
For a given number of observed S-stars, $N_{s},$ the parent external
population should be $1/f_{mig}$ larger to have $N_{par}=N_{s}/f_{mig}$
stars (compare with similar constraints derived by \citealp{per09}). 
Using our efficiency term, more ``efficient'' models are the ones having 
larger $f_{mig}$. 
Cases in which $f_{mig}$ is small, could be strongly constrained
by such requirement. For example, the 'billiard' model for the origin
of the S-stars \citep{ale+04} in which stars from the central pc
are captured close to the MBH through exchanges with SBHs close to
the MBH, is disfavored since the population of similar B-stars in
the central pc, are too small to accommodate the required parent population
\citep{pau+06}. Similarly, one can turn to models of a disk origin
for the S-stars, such as the Kozai-like perturbations in the two disks
scenario \citep{loc+09}, the eccentric disk instability model \citep{mad+09},
or the spiral density wave model \citep{gri10}. All of these models,
irrespective of details, suggest that the S-stars formed as part of
a stellar disk, up to a few 0.1 pc away from the MBH. A small fraction
of the stars formed in the disk migrated through some process (which works 
equally on stars of any mass),  to later
on become the currently observed S-stars. The current number of B-stars
inferred in the central 0.5, outside the central 0.04 pc where the
S-stars reside, is comparable, and likely somewhat smaller than the
total number of S-stars. These models are required to have a very
high migration efficiency in order to explain the origin of the S-stars.
Putting it differently, a much larger $f_{mig}$ than currently suggested
by these models is required. We can therefore conclude that (at least)
the current formulations of these models, can likely be excluded.

\subsection{External star-formation followed by rapid migration}

Another set of scenarios suggest that the GC young stars formed outside
the central pc, where conditions are less hostile for regular star
formation. These scenarios try to explain only the origin of the stellar
disk, or the origin of the isotropic B-stars, in particular the S-stars,
but not both. Note that the extension of S-stars distribution beyond
the central 0.04 pc is only a recent observational development, nevertheless,
some of the models suggested for the S-stars origin discussed the
possible existence of such extended distribution.

\subsubsection{Cluster infall}

An infall of a young stellar cluster (with or without an intermediate
MBH; IMBH) into the central pc was suggested as alternative scenario
for the origin of the GC young stars \citet{ger01,kim+03,por+03,kim+04,lev+05,gur+05,ber+06b,mak+07,fuj+09}.
Such dissolving cluster is likely to form a stellar disk like structure,
possibly with additional outlying structures and/or isolated stars
outside of the main disk, as observed in the GC. It would also produce
a bias towards more massive stars to be concentrated in the central
region of the GC. These more massive stars, which were originally
more segregated in the inner regions of the cluster, would be the
last to dissolve from the cluster, i.e. in the central most regions
closer to the MBH. 

Such inspiraling objects, however, may not be able to inspiral in
the appropriate time window for producing the stellar disk \citep{kim+03,kim+04}.
\citet{yuq+07} and others invoke the existence of an IMBH with mass
$>10^{4}\, M_{\odot}$ which could help produce the current observations.
Nevertheless, formation of such an object in a cluster and its infall
in the appropriate time window appears difficult \citep[e.g. ][]{kim+04}
and need to be fine tuned to explain the observations. In fact, detailed
N-body simulations of stellar clusters, combined with stellar evolution,
do not produce IMBH in these clusters \citep{gle+09,van+09}. The
possibility of an IMBH infall is therefore unfavorable given our currents
understanding and the lack of evidence for such objects in the central
region of the Galaxy.

Apart from the problems posed by requiring the existence of an IMBH,
cluster infall models may not be consistent with observations \citep{pau+06}.
An infalling cluster is likely to leave most of its stars behind during
its inpiral as it dissolves, where as very few young stars are observed
outside the central 0.5 pc. In a sense, this difficulty is similar to the
efficiency problem discussed previously for other models. 
Note that even the simulations of an infalling
cluster hosting an IMBH show that the young stars are stripped from
the cluster before the IMBH reaches the central 0.04 pc (see e.g.
\citealt{lev+05,ber+06b}). The young stars closest to the MBH (the
S-stars) are therefore not likely to directly originate from such
a scenario in any case.

\subsubsection{Binary disruption}

A close pass of a binary star near a MBH results in an exchange interaction,
in which one star is ejected at high velocity, while its companion
is captured by the MBH and is left bound to it \citep{hil88}. Such
interaction occurs because of the tidal forces exerted by the MBH
on the binary components. A young binary star could therefore be formed
outside the central region and later be scattered onto the MBH on
a highly radial orbit leading to its disruption. Such a scenario was
suggested by \citet{gou+03} to explain the origin of the star S2.
In order for the capture rate of such stars to explain the current
observation of all the GC B-stars, rapid relaxation processes are
required for the binaries to be scattered onto the MBH. Such a model,
suggested by \citet{per+07}, which takes into account scattering
by massive perturbers outside the central 1.5 pc (such as giant molecular
clouds and clumps observed in the GC region and other galactic nuclei;
\citealp{per+07,per+08}) could account for the observed number of
B-stars. Note that this scenario, like the disk origin models for
the S-stars discussed above, can be constrained by observations of
the parent population from which the S-stars originate. Current observations
do not exclude this model \citep{per+07,per09}. Future observation
looking for young stars in these regions should, in principle, give
better constraints. Nevertheless, other observational signatures may
be more easily verified or refuted. We discuss these in the following. 

The binary disruption scenario leaves the captured stars on highly
eccentric orbits (>0.94), and further dynamical evolution is required
in order to explain their currently observed eccentricity distribution.
Study of their evolution, which is driven  by resonant relaxation
processes \citep{rau+96}, suggest that indeed the more relaxed, almost
thermal eccentricity distribution of the S-stars \citep{per+09b}
could be consistent with their evolution from much higher initial
eccentricity. It is interesting to note, however, that the resonance
relaxation time scales in the GC increase with distance from the MBH
\citep{hop+06a}. Therefore stars captured further away from the MBH
are likely to have less relaxed eccentricity distribution, i.e. this
scenario produce a correlated eccentricity-distance, with stars more
distant from the MBH expected to have higher eccentricities. Stars
captured further than 0.5 pc, for example, are likely to be highly
eccentric (>0.94) in this scenario. 

Typically, a binary is disrupted when it crosses the tidal radius
of the MBH (see e.g. \citealp{per+09b} for details). One of the binary
components is typically captured at a close orbit near the MBH \citep{gou+03},
and its companion is ejected at high velocities \citep{hil88}. The
semi-major axis of the captured star around the MBH is linearly related
to its original binary progenitor separation \citep{hil91}. The radial
distribution of captured stars therefore maps the distribution of
the binaries separations. The distribution of semi-major axis of massive
binaries in the Solar-neighborhood follows a log-constant distribution.
The radial distribution of captured stars in the GC should therefore
be log-constant, or $n(r)\sim r^{-1}$, if the GC binaries have similar
properties. 

The mass function of captured stars is likely to be regular, i.e.
reflecting the mass function of stars far from the MBH, where regular
star-formation could occur. A small contribution from Kozai induced
merger of stars, induced by Kozai resonances near the MBH, may contribute
a small fraction of more massive stars \citep{ant+09,per09b,per+09c}.
In addition, longer living stars captured at earlier times may have
a higher probability of being disrupted by the MBH during their dynamical
evolution \citep{per+09b}. Taken together, the observed mass function
(MF) of captured B-stars is likely be quite regular, although possibly
more top heavy closer to the MBH, where disruption and merger occur,
than a regular MF expected for stars captured further away.

\section{Summary}
In these proceedings we have shortly reviewed the origin and evolution of the young stars in the Galactic center. These stars which could be divided into two distinct stellar populations likely originated from two different processes. The young stellar disk which contains mostly O and WR stars likely originated from an in-situ star formation through fragmentation of an infalling gaseous clamps. 
The population of young B-stars isotropically distributed throughout the central pc around the MBH likely have a different origin. Such stars were not likely to have been produced like the O-stars, and then migrated to their current postions, since a much larger parent population of B-stars should have been observed in the central pc. A binary disruption scenario, in which binaries which formed outside the central pc were scattered onto the MBH, could still be consistent with current observations. Such a scenario have specific predictions regarding the kinematic properties of the GC B-stars. We have reviewed these predictions which could be tested through direct observations in the coming few years.  
 
 



\end{document}